\documentclass[aps,prl,twocolumn,preprintnumbers,nofootinbib,showpacs]{revtex4}

\usepackage{graphicx}
\usepackage{amsmath}

\def\be{\begin{equation}}
\def\ee{\end{equation}}
\def\bea{\begin{eqnarray}}
\def\eea{\end{eqnarray}}

\def\afbt{A_{FB}^t}
\def\afbl{A_{FB}^\ell}
\def\coth{{\text{coth}}}

\begin{document}

\preprint{ANL-HEP-PR-11-90, MSUHEP-120109, IIT-CAPP-11-11}

\title{The Top Quark Production Asymmetries $A_{FB}^t$ and $A_{FB}^{\ell}$}

\author{Edmond L. Berger$^{b}$, Qing-Hong Cao$^{a}$, Chuan-Ren Chen$^{b}$,
Jiang-Hao Yu$^{c}$, Hao Zhang$^{b,d}$}
\affiliation{
\mbox{$^a$Department of Physics and State Key Laboratory of Nuclear Physics
and Technology, Peking University, Beijing, 100871, China}\\
\mbox{$^b$High Energy Division, Argonne National Laboratory, Argonne, IL 60439, USA}\\
\mbox{$^c$Department of Physics and Astronomy, Michigan State University, East Lansing, MI 48824, USA}\\
\mbox{$^d$Illinois Institute of Technology, Chicago, Illinois 60616-3793, USA}
}

\begin{abstract}
A large forward-backward asymmetry is seen in both the top quark rapidity distribution $A_{FB}^t$ 
and in the rapidity distribution of charged leptons $A_{FB}^\ell$ from top quarks produced at the 
Tevatron.  We study the kinematic and dynamic aspects of the relationship 
of the two observables arising from the spin correlation between the charged lepton and the 
top quark with different polarization states.   We emphasize the value of both measurements, and 
we conclude that a new physics model which produces more right-handed than left-handed top 
quarks is favored by the present data.
\end{abstract}


\maketitle

{\em Introduction.} The observed forward-backward asymmetry $A_{FB}^t$ in the rapidity distribution of top 
quarks~\cite{Aaltonen:2011kc, Abazov:2011rq} at the Tevatron deviates by about two standard deviations ($2\sigma$) from standard model (SM) expectations~\cite{Kuhn:1998jr}.   
In addition to $A_{FB}^t$, the D0 group also reports a positive forward-backward asymmetry of {\em charged leptons} from top quark decays of  $A_{FB}^\ell=(15.2\pm 4.0)\%$ compared with the small value $2.1\pm0.1\%$ from SM~\cite{Abazov:2011rq}.  
The deviation of the asymmetries may be contrasted with the good agreement of the overall rate for top quark production with SM predictions.  
 
In this Letter, we focus on the kinematic and dynamic relationship between $A_{FB}^t$ and $A_{FB}^\ell$.   
We investigate how the distribution of leptons in the laboratory frame is related to the polarization state of the top quark parent. 
We show in a model-independent manner that current data on the ratio of the two asymmetries imply that more right-handed than 
left-handed top quarks are produced.  This is a second and independent indication from asymmetry data of discrepancy from the SM since an equal number of right- and left-handed top quarks is predicted in the SM.  We urge confirmation of the D0 result by the CDF collaboration and with the full data set in D0.   Measurements of both $A_{FB}^t$ and $A_{FB}^\ell$ are especially valuable because their correlation can be related through top quark polarization to the underlying dynamics of top quark production.     

We begin with a discussion of the angular distribution of decay leptons, first in the rest frame of the top quark and then in the laboratory frame.     
Subsequently, we derive the relationship of $A_{FB}^\ell$ and $A_{FB}^t$ separately for  left- and right-handed top quarks.  Different models of new physics produce top quarks with different proportions of left- and right-handed polarization.  
We use a $W^{\prime}$ model~\cite{wprime} and an axigluon $G^{\prime}$ model~\cite{axi1} to deduce their different expectations for $A_{FB}^\ell/A_{FB}^t$.  The $W^\prime$ model and other models~\cite{Shu:2009xf} with more right- than left-handed top quarks tend to be preferred by the data provided that the constraint of the overall rate is satisfied. 

{\em Kinematics.} In the top quark rest frame, the distribution in the polar angle  $\theta_{\rm hel}$ of a decay lepton $\ell^+$ is~\cite{Mahlon:1995zn}
\be
\frac{1}{\Gamma}\frac{d\Gamma}{d\cos\theta_{\rm hel}}=\frac{1+\lambda_t\cos\theta_{\rm hel}}{2},
\label{eq:spin}
\ee
where $\lambda_t$ denotes the top quark helicity;  $\lambda_t=+$  for a right-handed 
($t_R$), and $\lambda_t=-$ for a left-handed top quark ($t_L$).  The angle is measured 
with resect to the direction of motion of the top quark in the laboratory frame.   
\begin{figure}
\includegraphics[scale=0.3]{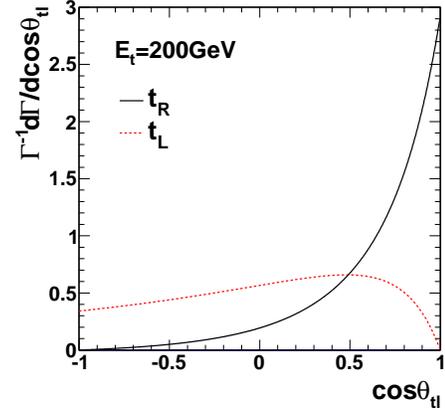}
\caption{
$\cos\theta_{t\ell}$ distribution in the boosted frame for a top quark 
with $E_t=200~{\rm GeV}$.
}
\label{fig:leprap1}
\end{figure}
\begin{figure}
\includegraphics[scale=0.2]{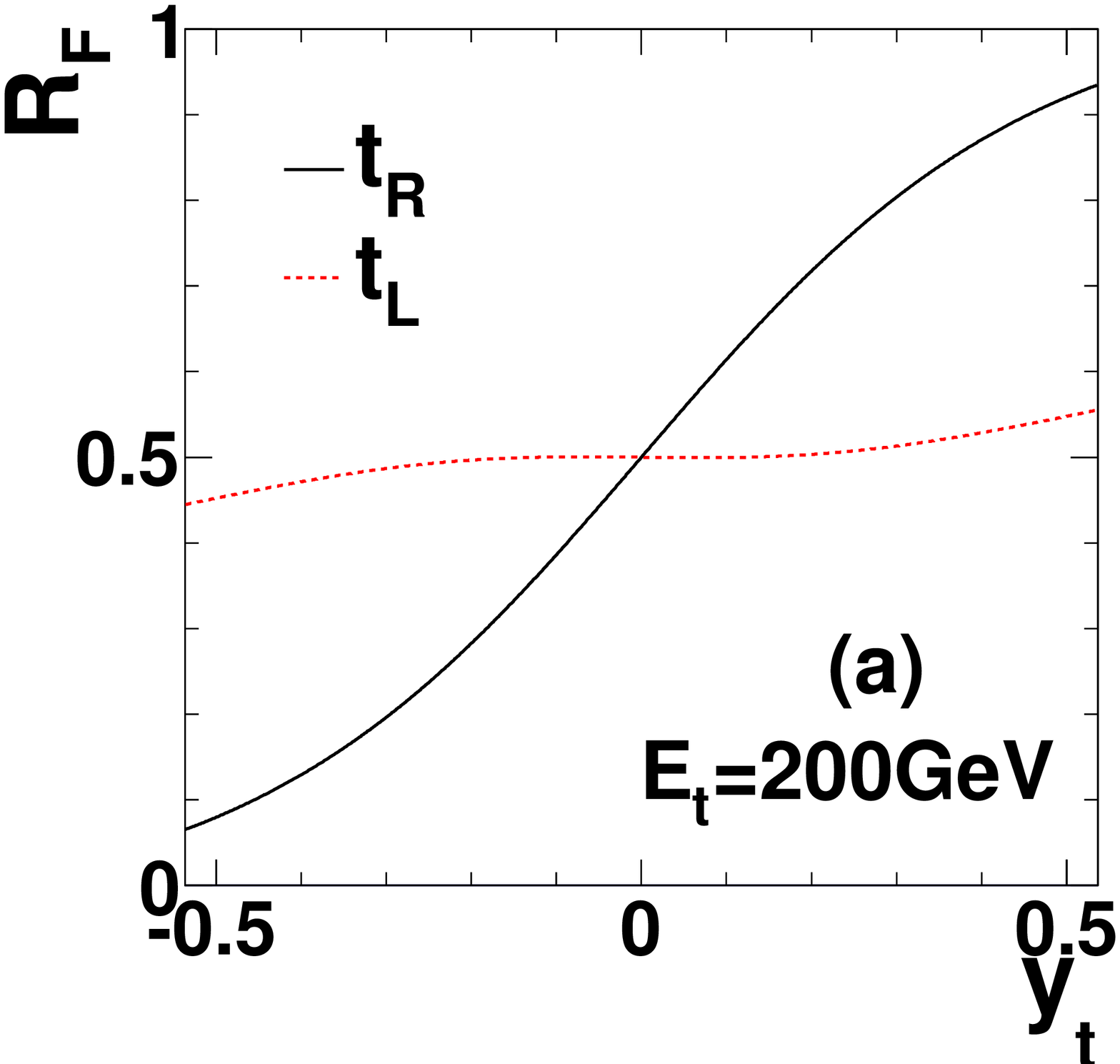}
\includegraphics[scale=0.2]{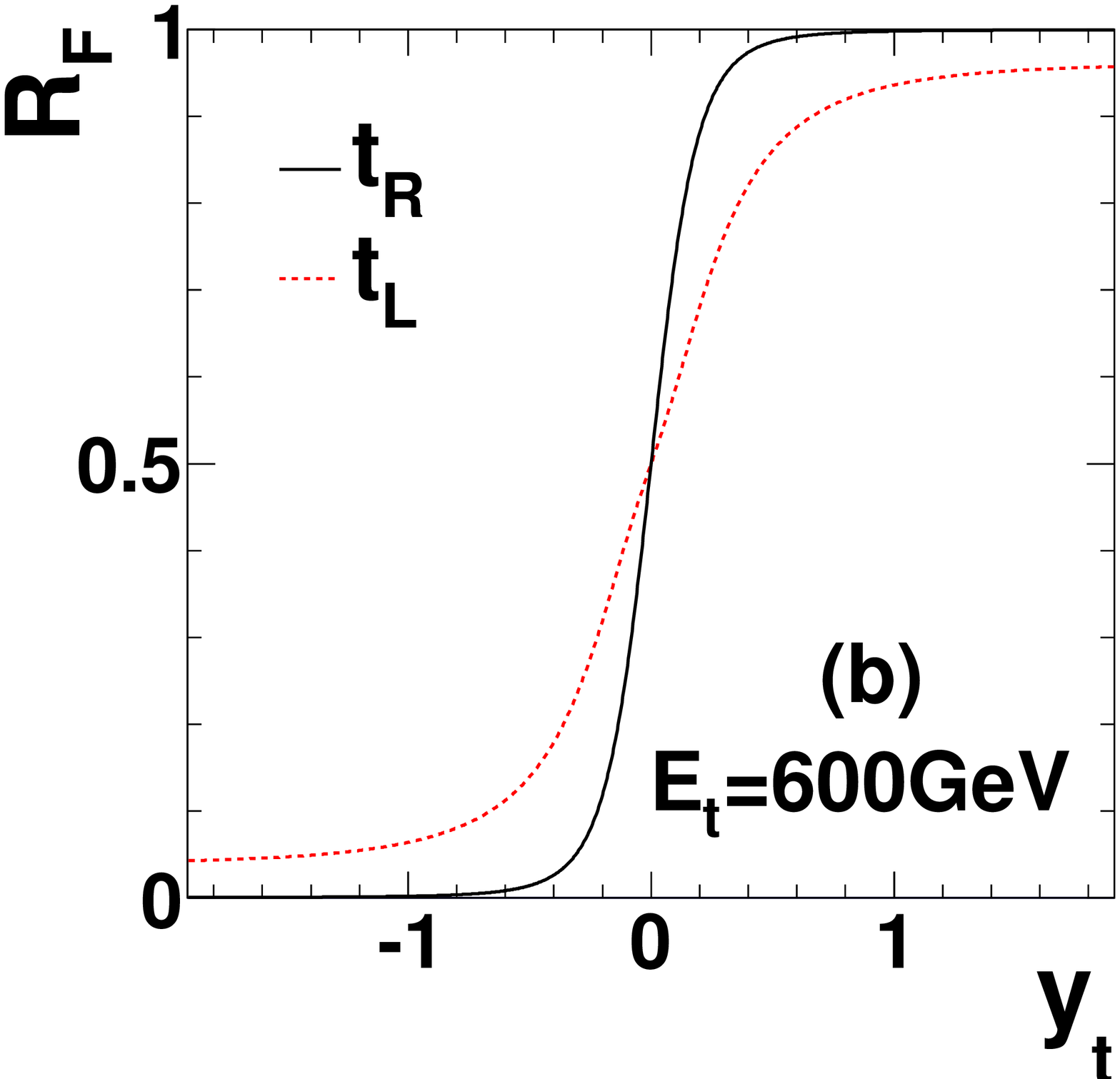}
\caption{
(a) The ratio $R_F$ as a function of $y_t$ for a top quark with $E_t = 200~{\rm GeV}$ and (b) $E_t=600~{\rm GeV}$.
}
\label{fig:leprap}
\end{figure}
Once the top quark is boosted, the angular distribution of the charged lepton relative to the 
direction of motion of the top quark is sensitive to 
the energy of the top quark $E_t$.  We derive    
\begin{equation}
\frac{d\Gamma}{\Gamma d\cos\theta_{t\ell}}=\frac{1-\beta\cos\theta_{t\ell}+\lambda_t\left(\cos\theta_{t\ell}-\beta\right)}{2\gamma^2\left(1-\beta\cos\theta_{t\ell}\right)^3}, \label{eq:lep_follow_top}
\end{equation}
where $\beta=\sqrt{1-m_t^2/E_t^2}$, $\gamma=E_t/m_t$, and $\theta_{t\ell}$ is the angle 
between $\ell^+$ and its parent top quark in the boosted frame.
As illustrated in Fig.~\ref{fig:leprap1}, for $E_t=200~{\rm GeV}$, 
about $60\%$ of $\ell^+$ follow the top quark (i.e., $\cos \theta_{t\ell} > 0$) for 
a $t_L$, and almost $100\%$ for a $t_R$. 

The top quark $y_t$ rapidity is $y_t \equiv \ln\sqrt{(E_t + p_z^t)/(E_t - p_z^t)}$ where 
$p_z^t$ is the longitudinal (z-component) of the top quark momentum.  The forward direction 
is specified as the direction of the incident proton beam.  The probability for finding a positive 
charged lepton in the forward region when it originates from a top quark with a velocity 
$\beta$, rapidity $y_t$, and polarization $\lambda_t$  is denoted 
 \begin{equation}
 R_F^{\ell,~\lambda_t}(\beta, y_t)=N_F^\ell/\left(N_F^\ell+N_B^\ell\right),
\end{equation}
where $N_F^\ell$ ($N_B^\ell$) is the number of leptons $\ell^+$ in the 
forward (backward) region in the laboratory. 
After lengthy algebra, we derive 
\begin{eqnarray}
R_F^{\ell,\lambda_t}(\beta, y_t)&=&
\displaystyle \frac{1}{2}+\frac{1}{2\left(1+\gamma^{-2}\coth^2y_t\right)^{1/2}}\nonumber\\
&&+\frac{\lambda_t\coth^2y_t}{4\beta\gamma^{2}\left(1+\gamma^{-2}\coth^2y_t\right)^{3/2}}
\end{eqnarray}
for $y_t\in\left[0,y_t^{\rm{max}}\right]$, where $y_t^{\rm max}=\ln\sqrt{\left(1+\beta\right)/\left(1-\beta\right)}$.
 
To illustrate the effect of the top quark boost, we plot  
$R_F$ as a function of $y_t$ in Fig.~\ref{fig:leprap}(a,b).  We choose two characteristic top quark energies,
$E_t=200~{\rm GeV}$ and 600~GeV. 
The former energy represents top quarks produced around the threshold region, while the latter 
pertains for highly boosted top quarks. 
When a top quark moves perpendicular to the beam line, i.e. $y_t=0$, there is an equal 
number of leptons in the forward and backward regions, i.e. $R_F=0.5$, independent 
of $E_t$ and $\lambda_t$.  

For $t_R$, $R_F$ increases rapidly with $y_t$
because most of the leptons move close to the direction of motion of the top quark after being 
boosted to the lab frame.   We 
can also see that when $E_t$ becomes larger, i.e. the top quark is more energetic and the lepton 
is more boosted, $R_F$ rapidly reaches its maximum value $1$. 

On the contrary, in the case of $t_L$'s, the ratio $R_F$ does not vary significantly with $y_t$ owing 
to the anti-boost effect on $\ell^+$.  For $E_t = 200$ GeV, the boost causes $\ell^+$
to distribute nearly uniformly, and $R_F$ is around $0.5$.   When the energy of $t_L$'s is large enough, the large boost forces most 
of the charged leptons from top quark decays to move along the top quark direction of motion, even 
if they move against the top quark direction of motion in the top quark rest frame.  The boost yields a 
large value $R_F$ in the region of large $y_t$.  The competing influences leave the $t_L$ curve slightly below the $t_R$ curve.  
 
\begin{figure}
\includegraphics[scale=0.21]{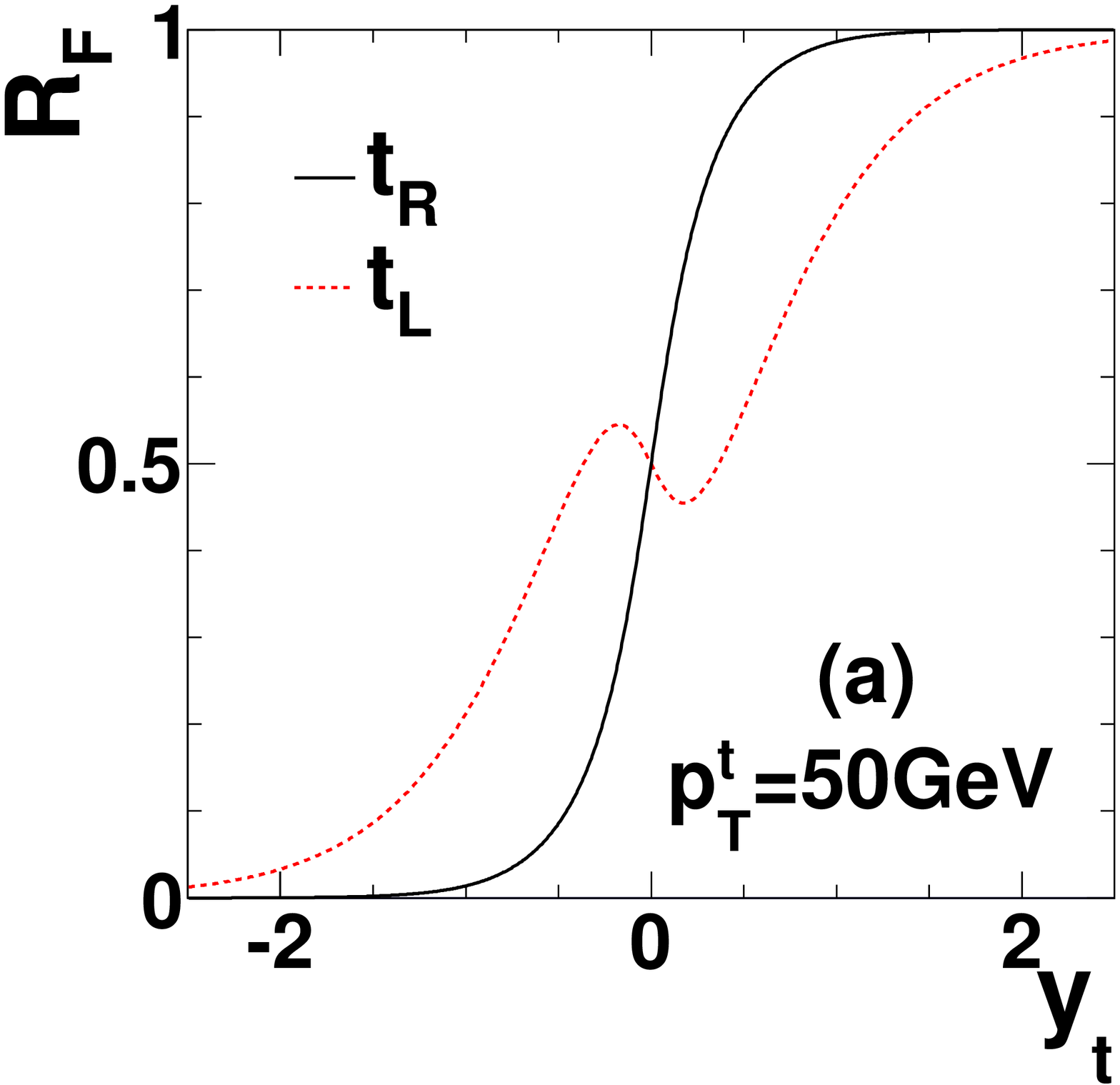}
\includegraphics[scale=0.21]{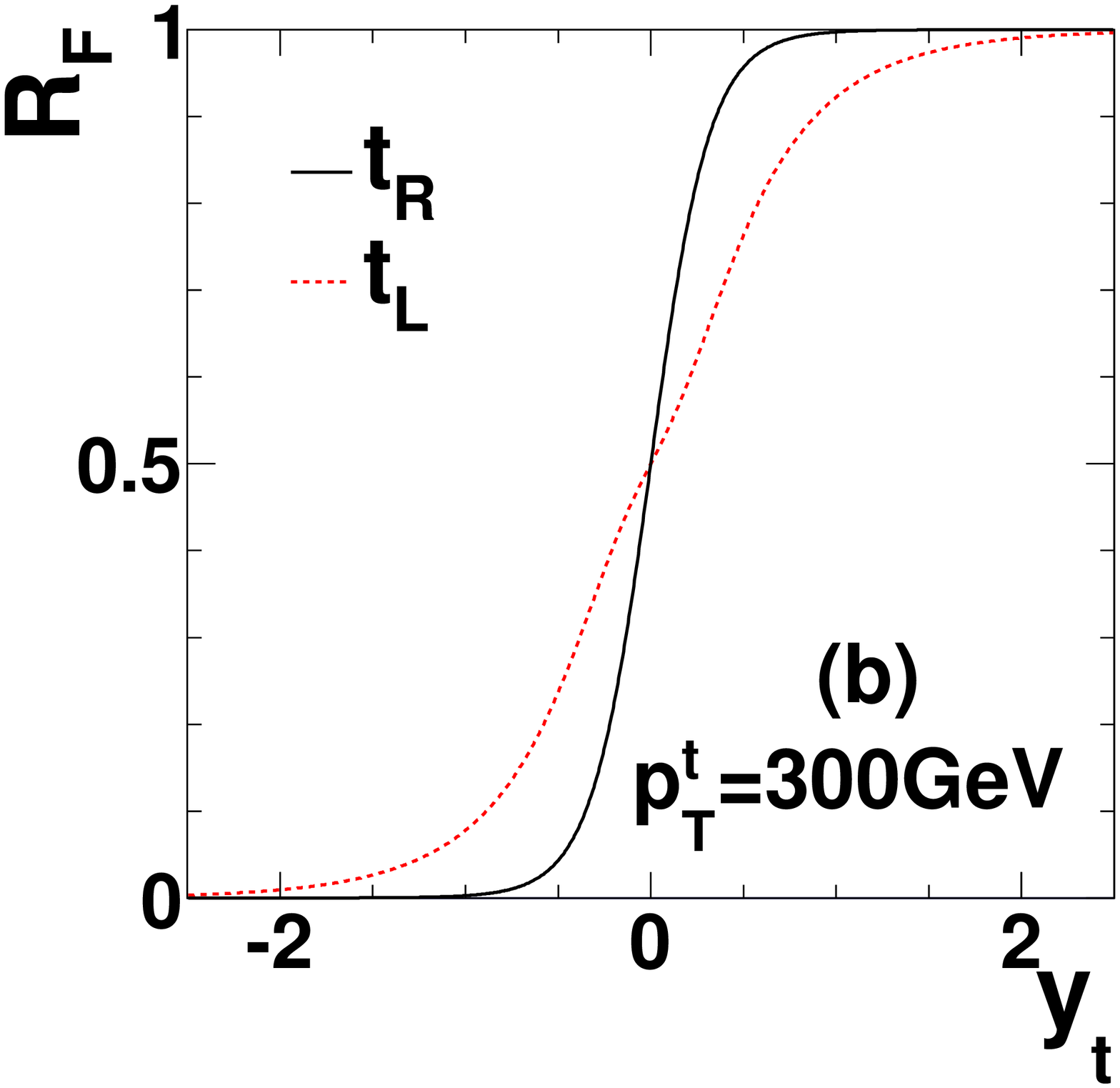}
\caption{(a) $R_F$ as a function of $y_t$
for  top quarks with fixed $p_{T}^t =50~{\rm GeV}$ and (b) $p_{T}^t = 300~{\rm GeV}$.} 
\label{fig:leprap4}
\end{figure}

In Fig.~\ref{fig:leprap4}, we show how  $R_F$ varies with $p_T^t$ and $y_t$.  The distributions for $t_R$'s do not vary greatly with $p_T^t$ because most $\ell^+$ follow $t_R$.   However, 
the shapes of the curves for $t_L$'s are very 
different between the low $p_T^t$ and high $p_T^t$ regions.  As the top 
quark moves forward, i.e. $y_t > 0$ for fixed $p_T^t$, the boost becomes more significant as the energy of the 
top quark is increased.  Therefore, more leptons are forced to move along the direction of the top quark.  On 
the other hand, some fraction of the decay leptons which are initially in the forward/backward region 
($y_\ell > 0/y_\ell <0$) will then be in the backward/forward region. 
In summary, two factors affect $R_F$: the boost and the rearrangement of the distribution of 
charged leptons in the forward ($y_\ell>0$) and backward ($y_\ell<0$) regions. 
The former always increases $R_F$ while the latter may increase or decrease  $R_F$ depending on $E_t$ at $y_t =0$.  
Generally speaking, when the initial boost is not significant (low $p_T^t$), $R_F$ decreases when 
$y_t$ increases from $y_t=0$, as we see in Fig.~\ref{fig:leprap4}(a).  For large enough boost 
($p_T^t>m_t/ \sqrt{3}$), $R_F$ always increases with $y_t$;  the critical value is obtained from 
$\frac{\partial R_F}{\partial y_t}|_{y_t=0} = 0$.  

{\em $A_{FB}^t$ and $A_{FB}^\ell$.  } Positive $A_{FB}^t$ indicates more top quarks are produced  in 
the forward region than in the backward region.
Both $t_R$ and $t_L$ can generate a positive $A_{FB}^\ell$.  
However, $t_L$ would need a large boost along the beam 
line to overcome the fact that most of $\ell^+$
from its decay move against it in its rest frame, while 
$t_R$ can yield a positive $A_{FB}^\ell$ even for top quarks near 
the $t\bar{t}$ threshold region.  Therefore, the observed positive  
$A_{FB}^t$ and $A_{FB}^\ell$ indicate that
the top quark polarization may be playing a non-trivial role.   In this section we present a general  
analysis of the correlation between $A_{FB}^t$ and $A_{FB}^\ell$, to prepare for a better 
understanding of the numerical results derived from  new physics (NP) models.

Assuming the large $A_{FB}^t$ is generated mainly by NP,
 $A_{FB}^t$ can be divided into the contributions from different polarizations of top quarks: 
\be
A_{FB}^t \approx\left[\rho_{t_L}~A_{FB}^{t_L,~{\rm NP}} +\rho_{t_R}~A_{FB}^{t_R,~\rm NP}\right]\times R^{\rm NP},
\ee
where
\be
A_{FB}^{\lambda_t,~\rm NP}=\left[\frac{N_F^{\lambda_t}-N_B^{\lambda_t}}
{N_F^{\lambda_t}+N_B^{\lambda_t}}\right]_{\rm  NP}, \quad
\rho_{\lambda_t} = \frac{N^{\lambda_t,~\rm NP}}
{N_{\rm tot}^{\rm NP}}.
\ee
Here, $A_{FB}^{\lambda_t,~\rm NP}$ denotes the forward-backward asymmetry of the top quark
with polarization $\lambda_t$ generated only by NP, while $\rho_{\lambda_t}$ is
the fraction of  top quarks with polarization $\lambda_t$ in $t\bar{t}$ events induced 
by NP, and $R^{\rm NP}(=N^{\rm NP}_{\rm tot}/N_{\rm tot}$) is the ratio of NP signal events to 
the total observed $t\bar{t}$ events.
One advantage of decomposing $A_{FB}^t$ into different top quark polarizations
is to monitor the chirality of the couplings of NP particles to top quarks. Another advantage is to 
make the connection between $A_{FB}^\ell$ and $A_{FB}^t$ more transparent. 

As discussed earlier, the ratio $R_F^\ell$ depends on the top quark kinematics 
($\beta$, $y_t$ and $\lambda_t$).   To compute the probability for a charged lepton in the 
forward region,  one must convolute the top quark production cross section with $R_F^\ell$ 
on an event-by-event basis, i.e. 
\be
N^{t\bar{t}}\otimes R_F^{\ell,\lambda_t} = \int N^{t\bar{t}}(\beta, y_t, \lambda_t) R_F^{\ell,\lambda_t} (\beta, y_t),
\ee
where $N^{t\bar{t}}$ labels the $t\bar{t}$ production rate for a top quark with specific kinematics
($\beta$, $y_t$, $\lambda_t$).  
The lepton asymmetry $A_{FB}^{{\ell}, \lambda_t}$ generated by a top quark with 
polarization $\lambda_t$ is, therefore, 
\bea
A_{FB}^{\ell,\lambda_t}\bigg|_{\rm NP} &=& \left.\frac{(N_F^{\lambda_t}-N_B^{\lambda_t})\otimes
\left(2 R_F^{\ell,\lambda_t}-1\right)}{N_F^{\lambda_t}+N_B^{\lambda_t}}\right|_{\rm NP} .
\label{eq:correlation}
\eea 
Because $R_F^{\ell,\lambda_t}$ cannot exceed 1, we have $A_{FB}^\ell \lesssim A_{FB}^t$.
When $R_F^{\ell, \lambda_t}$ is close to a constant $\mathcal{R}_C$, 
e.g. $\mathcal{R}_C \sim 1/2$ around the $t\bar{t}$ threshold ($E_t\sim200{\rm GeV}$) for left-handed top quark
or $\mathcal{R}_C\sim 1$ for a highly boosted top quark, 
the lepton asymmetry $A_{FB}^{\ell, \lambda_t}$ 
can be simplified as
\bea
A_{FB}^{\ell, \lambda_t}\big|_{\rm NP} =A_{FB}^{\lambda_t,~{\rm NP}} \times  \left(2\mathcal{R}_C-1\right).
\label{eq:correlation2}
\eea
Equation~(\ref{eq:correlation2}) and Fig.~\ref{fig:leprap} show that:
\begin{itemize}
\item $A_{FB}^{\ell, t_L} \sim 0$ when the $t\bar{t}$ pair is produced around the threshold region;
\item $A_{FB}^{\ell, t_L} \lesssim A_{FB}^{\ell, t_R} \approx A_{FB}^{t}$ in the large $m_{t\bar{t}}$ region.
\end{itemize}
Although Eq.~(\ref{eq:correlation2}) is approximate, it helps in understanding the NP prediction obtained 
from a complete numerical calculation.

{\em New physics models: axigluon and $W^\prime$}.  We examine two models of new physics, an axigluon 
model~\cite{axi1} 
and a flavor-changing $W^\prime$ model~\cite{wprime}. In the axigluon ($G^\prime$) model we assume for simplicity that the interaction of 
$G^\prime$ to the SM quarks is purely pseudo-vector-like
\be
\mathcal{L} = g_s \left(g_l~\bar{q}\gamma^\mu \gamma_5 q 
+ g_h~\bar{Q}\gamma^\mu \gamma_5 Q\right) G^{\prime}_{\mu},
\ee
where $q$ denotes the first two generation quarks and $Q$ the third generation
quarks.   The coupling $g_s$ is the strong coupling strength; $g_l$ and $g_h$ are the 
coupling strength of $G^\prime$ to 
$q$ and $Q$, respectively.

The absence of deviation from the SM expectation in the measured $m_{t\bar{t}}$ 
distribution~\cite{Aaltonen:2011kc, Abazov:2011rq}
indicates the $G^\prime$ should be heavy and broad.  Its contribution is therefore 
through interference with the SM channel.  
\begin{figure}
\includegraphics[scale=0.27]{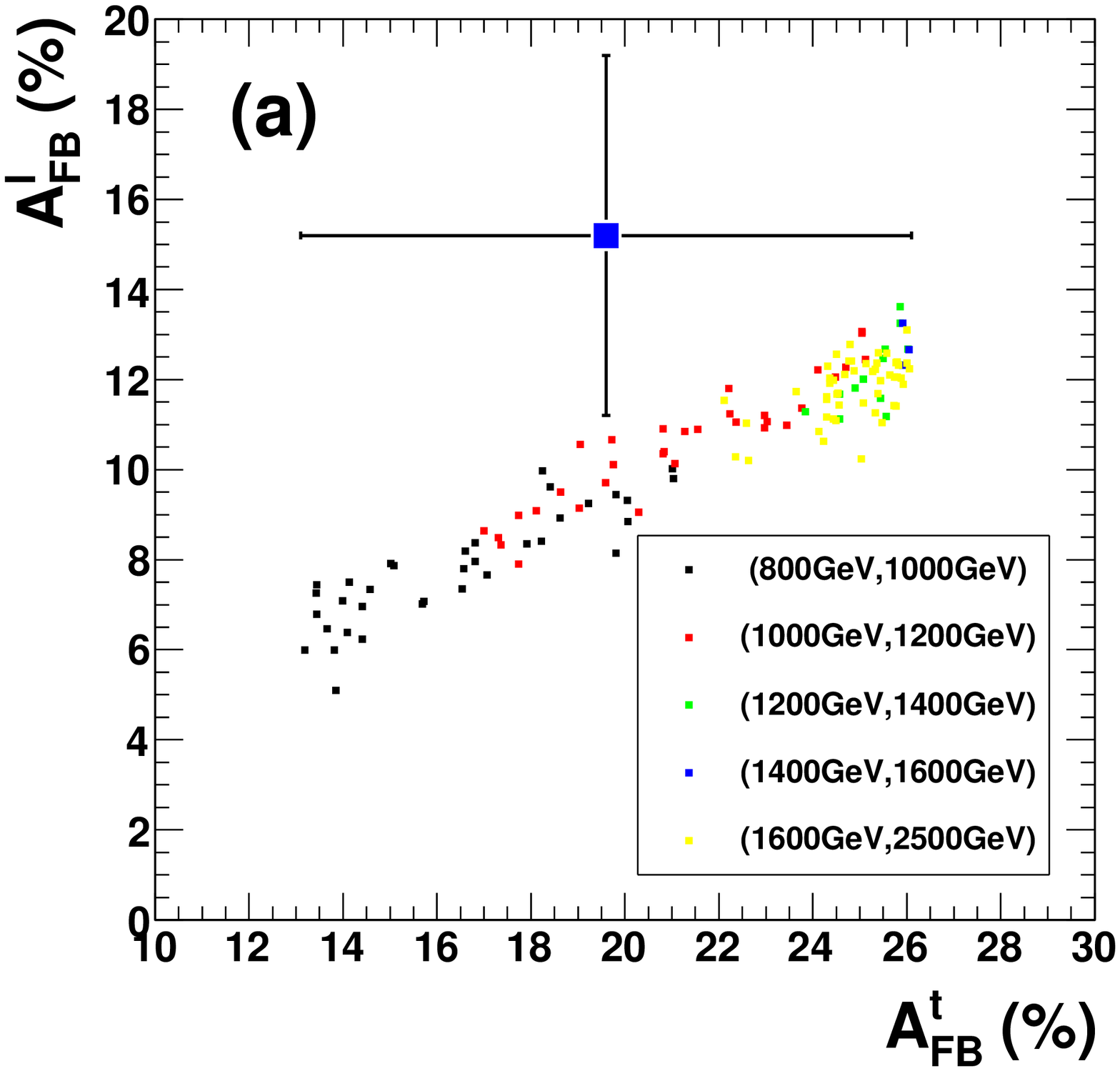}
\includegraphics[scale=0.27]{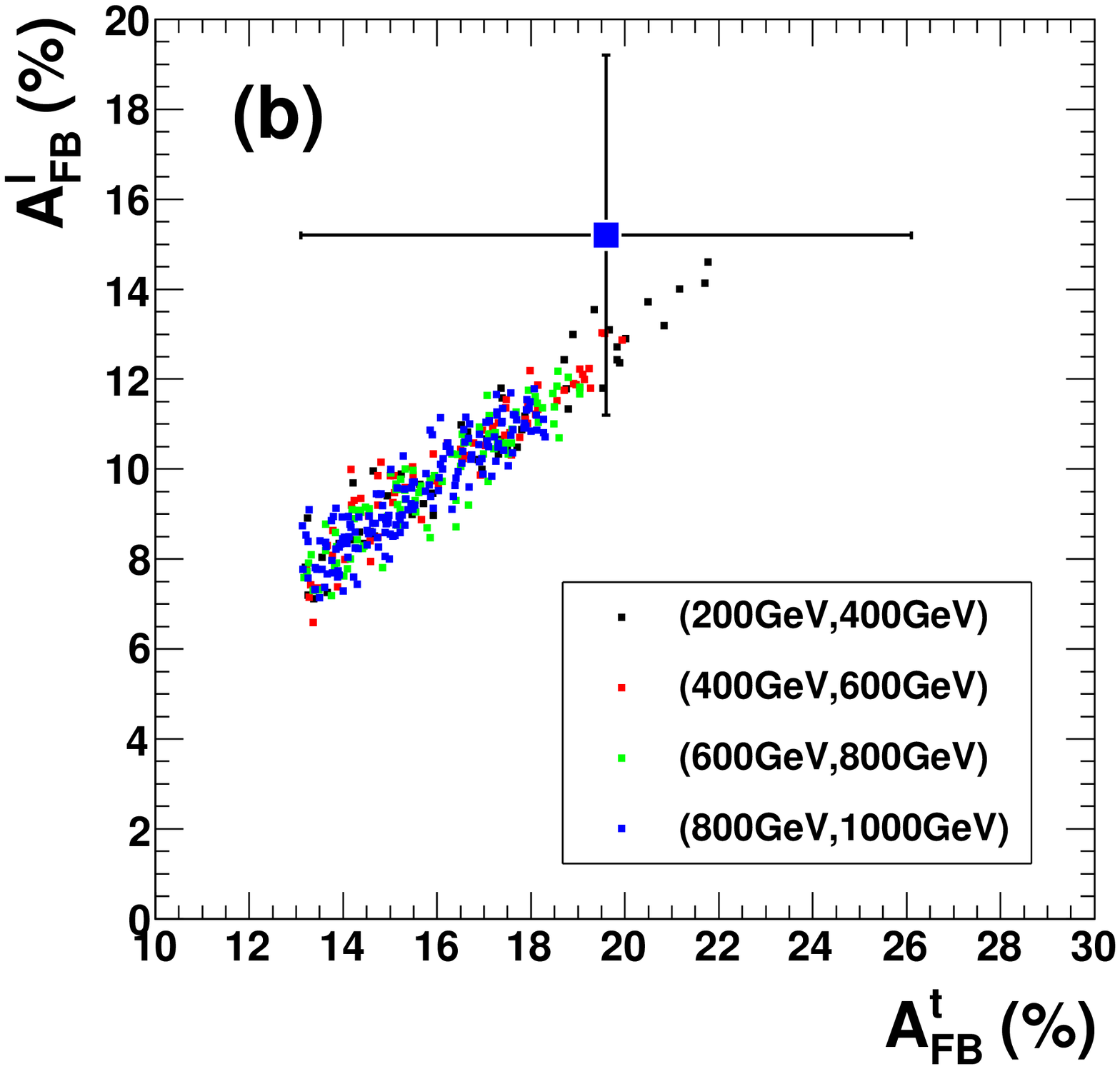}
\caption{Correlation between $A_{FB}^{\ell}$ and $A_{FB}^{t}$ for (a) the axigluon
and (b) the $W^\prime$ models.  The point corresponding to the D0 data is also shown.   
The numbers within the parentheses label the lower and upper limits of the mass of the 
NP  object.   
For comparison, the SM values are 
$A_{FB}^t \sim 5\%$ (off the left side of the plots in (a) and (b)), and 
$A_{FB}^\ell \sim 2\%$. }
\label{fig:correlation}
\end{figure}
The top quarks are generated unpolarized owing to the pseudo-vector coupling of 
the $G^\prime$ to the SM fermions, and
\be  
\rho_{t_L}=\rho_{t_R}=1/2, A_{FB}^{t_L,~{\rm NP}}=A_{FB}^{t_R,~{\rm NP}}=A_{FB}^t/R^{\rm NP}>0.
\ee
Since the $t \bar{t}$ cross section is greatest near the threshold region where 
$A_{FB}^{\ell, t_L} \sim 0$ and $A_{FB}^{\ell, t_R}\sim A_{FB}^t$, the expression for 
$A_{FB}^\ell$ becomes $A_{FB}^{\ell} \sim  \frac{1}{2} A_{FB}^t$.

We plot our axigluon model predictions for $A_{FB}^t$ and $A_{FB}^{\ell}$ 
in Fig.~\ref{fig:correlation}(a).
We first scan the theoretical parameter space ($g_l$, $g_h$ and $m_{G^\prime}$) 
to fit Tevatron data on $A_{FB}^t$ 
and the $t\bar{t}$ total production cross section within  $1~\sigma$. 
These parameters are then used to calculate $A_{FB}^\ell$.  
The figure shows a clear correlation between $\afbt$ and $\afbl$.  The best fit to the 
correlation is $A_{FB}^\ell \simeq 0.47 \times A_{FB}^t + 0.25\%$. 
To fit both $A_{FB}^t$ and $A_{FB}^\ell$ within $1\sigma$, the mass of the $G^\prime$ 
must be greater than $1$~TeV.  For masses this great,  top quarks from $G^\prime$ decays 
are highly boosted and cause more $\ell^+$ to move along the direction of 
the top quarks.  
We remark here that if the $G^\prime$ is found as a resonance in the $t{\bar t}$ 
mass distribution, the chirality structure of its coupling to $t{\bar t}$ can possibly be determined 
at the LHC~\cite{Berger:2011hn}.
 
A different class of NP models to explain the $A^t_{FB}$ is based 
on $t$-channel kinematics.  
A model with a non-universal massive neutral vector boson $Z^\prime$~\cite{Jung:2009jz} is disfavored because it implies an excessive rate for same-sign 
top quark production at the 7~TeV LHC~\cite{Berger:2011ua}.

We consider in this paper a flavor-changing $W^\prime$ which couples an incident $d$-quark to the 
produced $t$-quark~\cite{wprime},
\be
\mathcal{L}=g_2 g_R \bar{d}\gamma^\mu P_R t W^\prime_{\mu} + h.c.~, 
\ee
where $g_2$ is the weak coupling.
In the $W^\prime$ model, in addition to the SM process 
the $t\bar{t}$ pair can also be produced via a $t$-channel process with
a $W^\prime$ mediator.  
In the  region $\beta\simeq 1$, the nonzero helicity 
amplitudes $\mathcal{M}_{W^\prime}^{t}(\lambda_q,\lambda_{\bar q}, \lambda_{t},\lambda_{\bar t})$ are
\bea
\mathcal{M}_{W^\prime}^t(+--+)&\sim&2r_W^2(1-\cos\theta), \nonumber \\
\mathcal{M}_{W^\prime}^t(+-+-)&\sim&4(1+\cos\theta)~
\eea
where $r_W=m_t/m_{W^\prime}$.
In order to produce top quarks in the forward region, one needs 
$2r_W^2<4$, which is always true for the region of $W^\prime$ masses (heavier than the top quark) considered in this paper.  
At the Tevatron the $\beta$ distribution of the top quark in $t\bar{t}$ production peaks around $0.6$, and therefore most of the 
top quarks are not significantly boosted.  We can also easily see that $\rho_{t_R} > \rho_{t_L}$ in the $W^\prime$ model.
Since the $t$-channel propagator contributes a minus sign,  $A_{FB}^t$ arises from 
a competition between the square of the purely NP term and the interference term of NP with the SM.  
The strong correlation is fit well 
by $A_{FB}^\ell \simeq 0.75 \times A_{FB}^t - 2.1\%$. 
Moreover, for a relatively light $W^\prime$ ($\lesssim 600$) GeV, both $\afbt$ and $\afbl$ can be 
consistent with the D0 data within $1~\sigma$. 

The ratio of the 
predicted $\afbl$ to $\afbt$
peaks near $50\%$ in the axigluon model and near $62\%$ in the $W^\prime $ model. The data from D0 shows about  $78\pm 33\%$.
The ratio in the SM is close to $40\%$. 
The $W^\prime$ model generates a larger $\afbl$ than the axigluon $G^\prime$ model because it produces 
more right-handed top quarks.  The comparison to the D0 point shown in Figs.~\ref{fig:correlation}(a,b) indicates 
that top quark events with a large proportion of right-handed top quarks are favored.   Constraints on flavor-changing 
currents in the $W^\prime$ model allow only right-handed couplings to the top quark, consistent with the D0 $\afbl$ 
results.   There is no direct evidence of the handedness of the coupling in the massive gluon models.  The D0 result 
could be interpreted as an indirect clue for the chiral couplings of the massive gluon.  

{\em Summary.} 
We study the kinematic and dynamic aspects 
of the relationship between the asymmetries $\afbt$ and $\afbl$ based on the spin correlation between charged leptons 
and the top quark with different polarization states.   Owing to the spin correlation in top quark decay, 
$\afbl$ and $\afbt$ are strongly positively correlated for {\em right-handed} top quarks.   However, for {\em left-handed} 
top quarks, the nature of the correlation depends on how boosted the top quark is.   For large enough $E_t$, $t_L$ will also generate a large $A_{FB}^\ell$, similar to that for $t_R$.  However, if $t_L$ is not boosted, $\afbl$ from it will be less than $\afbt/2$ for a 
positive $\afbt$.  Since most of the $t\bar{t}$ events are  produced in the threshold region, one may use the large positive 
values of $\afbt$ and $\afbl$ measured at D0 to conclude that production of left-handed top quarks is disfavored.   
Confirmation of the D0 result and greater statistics are essential.  There is great value in making measurements of 
both $A_{FB}^t$ and $A_{FB}^\ell$ because their correlation can be related through top quark polarization to the underlying 
dynamics of top quark production.

\begin{acknowledgments}~The work of E.L.B., C.R.C. and H.Z. is supported in part by the U.S.
DOE under Grants No.~DE-AC02-06CH11357. H.Z. is also supported by DOE under the Grant No. DE-FG02-94ER40840.    
The work of J.H.Y. is supported in part by the U.S. National Science Foundation 
under Grand No. PHY-0855561.
\end{acknowledgments}


\begin{thebibliography}{99}


\bibitem{Aaltonen:2011kc}
  T.~Aaltonen {\it et al.}  [CDF Collaboration],
  Phys.\ Rev.\  D {\bf 83}, 112003 (2011).

\bibitem{Abazov:2011rq}
  V.~M.~Abazov {\it et al.}  [D0 Collaboration], Phys.\ Rev.\  D {\bf 84}, 112005 (2011).

\bibitem{Kuhn:1998jr}
  J.~H. K{\"u}hn and G.~Rodrigo, Phys. Rev. Lett. {\bf 81}, 49 (1998); 
Phys.\ Rev.\  D {\bf 59}, 054017 (1999).
  
\bibitem{wprime}
K.~Cheung, W.~Y.~Keung and T.~C.~Yuan, Phys.\ Lett.\  B {\bf 682}, 287 (2009).

\bibitem{axi1}
P.~Ferrario and G.~Rodrigo, Phys.\ Rev.\  D {\bf 80} (2009) 051701;
 P.~H.~Frampton, J.~Shu and K.~Wang, Phys.\ Lett.\  B {\bf 683}, 294 (2010);
  Q.-H. Cao, D.~McKeen, J.~L. Rosner, G.~Shaughnessy, and C.~E.~M. Wagner, Phys. Rev. {\bf D81}, 114004 (2010).

\bibitem{Shu:2009xf} 
  J.~Shu, T.~M.~P.~Tait and K.~Wang,
  Phys.\ Rev.\ D {\bf 81}, 034012 (2010);
  A.~Djouadi, G.~Moreau and F.~Richard,
  Phys.\ Lett.\  B {\bf 701}, 458 (2011);
  D.~Krohn, T.~Liu, J.~Shelton and L.~T.~Wang,
  arXiv:1105.3743 [hep-ph];
  A.~Falkowski, G.~Perez and M.~Schmaltz,
  arXiv:1110.3796 [hep-ph].
  
\bibitem{Mahlon:1995zn}
  G.~Mahlon and S.~J.~Parke,
  Phys.\ Rev.\  D {\bf 53}, 4886 (1996).

\bibitem{Berger:2011hn}
  E.~L.~Berger, Q.~H.~Cao, C.~R.~Chen and H.~Zhang,
  Phys.\ Rev.\  D {\bf 83}, 114026 (2011).

\bibitem{Jung:2009jz}
  S.~Jung, H.~Murayama, A.~Pierce, and J.~D. Wells, Phys. Rev. {\bf D81}, 015004 (2010).

\bibitem{Berger:2011ua}
  E.~L.~Berger, Q.~H.~Cao, C.~R.~Chen, C.~S.~Li and H.~Zhang,
  Phys.\ Rev.\ Lett.\  {\bf 106} (2011) 201801;
   S.~Chatrchyan {\it et al.}  [CMS Collaboration],
  JHEP {\bf 1108} (2011) 005
    

\end{thebibliography}
\end{document}